\documentclass[aps,prb,amsmath,amssymb,twocolumn,superscriptaddress,showpacs,floatfix]{revtex4-2}

\usepackage{graphicx}
\usepackage{dcolumn}
\usepackage{bm}
\usepackage{amsmath}
\usepackage{amssymb}
\usepackage{multirow}

\begin{document}

\title{Extrinsic tunnel Hall effect in CoFeB/MgO/Pt junctions}

\author{I.~Yu.~Pashenkin}
\affiliation{Institute for Physics of Microstructures RAS, Nizhny Novgorod, 603950, Russia}

\author{M.~V.~Sapozhnikov}
\affiliation{Institute for Physics of Microstructures RAS, Nizhny Novgorod, 603950, Russia}
\affiliation{Lobachevsky State University of Nizhny Novgorod, 603950 Nizhny Novgorod, Russia}

\author{N.~S.~Gusev}
\affiliation{Institute for Physics of Microstructures RAS, Nizhny Novgorod, 603950, Russia}
\affiliation{Lobachevsky State University of Nizhny Novgorod, 603950 Nizhny Novgorod, Russia}

\author{E.~A.~Karashtin}
\affiliation{Institute for Physics of Microstructures RAS, Nizhny Novgorod, 603950, Russia}
\affiliation{Lobachevsky State University of Nizhny Novgorod, 603950 Nizhny Novgorod, Russia}

\author{A.~A.~Fraerman}
\affiliation{Institute for Physics of Microstructures RAS, Nizhny Novgorod, 603950, Russia}

\date{\today}


\begin{abstract}
The Hall effect that occurs when current flows through a CoFeB/MgO/Pt tunnel junction is investigated. It is shown that the transverse voltage in Pt electrode is nonlinear on a DC voltage applied to the tunnel junction. It has both linear (odd) and quadratic (even) parts. The linear part contains well-known contributions of the anomalous Hall effect in the ferromagnetic electrode, inverse spin-hall effect in platinum and others. The quadratic part is a phenomenon caused by the spin-orbit scattering of electrons in an external electric field induced by a voltage applied to the barrier. This field reaches values of $10^9$~V/m which is close to internal atomic fields. The magnitude of both effects decreases as thickness of Pt electrode is increased due to shunting effects.
\end{abstract}

\maketitle

The spin-orbit coupling (SOC), which is a relativistic contribution to the interaction between a particle and an electric field, is characterized by the energy
\begin{equation}
H_{SO}=\lambda \bm{\sigma}\cdot\mathbf{E}\times\mathbf{p}
\end{equation}
which combines the electric field $\mathbf{E}$, particle spin $\boldsymbol{\sigma}$ and momentum $\mathbf{p}$. Despite the fact that it is a relativistic interaction, the effects associated with it appear in a significant way especially in the solid states where SOC is enhanced by a factor of about $m_0c^2/E_G$, where $E_G\sim1$~eV is the gap~[\onlinecite{Rashba,Petersen}]. There spin-orbit effects are mainly caused by the motion of electrons, which act as charge and spin carriers, and the electric field of atomic nuclei. That is why SOC is strongest in metals with a large atomic weight, such as Pt, Ta, W and Bi. SOC can affect both localized electrons and conduction electrons. In the first case it leads to well-known phenomena of magnetic anisotropy. In the case of electron transport, SOC results to spin Hall effect (SHE)~[\onlinecite{Sinova}]. The effect consists of the appearance of spin accumulations on the lateral surfaces of an electric current-carrying sample. It was originally predicted by Dyakonov and Perel in 1974~[\onlinecite{Dyakonov-1,Dyakonov-2}] and then discovered experimentally in 2004~[\onlinecite{Kato}]. The inverse SHE (ISHE) is observed when a pure spin current (or a spin polarized electric current) flows in a material with a strong SOC. This leads to the appearance of a transverse charge current (or to charge accumulation on the lateral surfaces)~[\onlinecite{Averkiev,Bakun}]. A particular example of ISHE is the anomalous Hall effect (AHE) since the electric current in magnetic materials is initially spin-polarized due to the internal exchange splitting.

SHE and ISHE has now become the basis of a rapidly developing new direction of condensed state physics - spin orbitronics~[\onlinecite{Liu-1}]. It is believed that a possibility of mutual conversion of electric and spin currents will open the prospect for energy-efficient logic and memory devices for information processing~[\onlinecite{Liu-1,Kuschel,Manchon,Manipatruni,Dowben}]. 

In transition metals such as Co or Fe, which are used as spin-polarized current injectors, the degree of spin polarization of electrons is about 60~$\%$~[\onlinecite{Meservey}]. The use of a tunnel magnetic junction with a MgO dielectric layer together with CoFeB ferromagnet as an injector makes it possible to increase the spin polarization to almost 100~$\%$~[\onlinecite{Zhang}]. So SHE in the systems with tunnel barrier has also been studied recently~[\onlinecite{Liu-2,Liu-3,Fang,Nakagawara,Gotte}]. In this regard, a question arises about the possible effects of SOC directly in the tunnel gap. Scattering by magnetic impurities in the barrier and a frozen-in boundary electric field are considered in~[\onlinecite{Vedyayev-1}] and~[\onlinecite{Tarasenko,Matos}], respectively.
In all above examples the electric field causing SOC according to (1) is the internal atomic field. In~[\onlinecite{Vedyayev-2}] they consider theoretically the problem of the SOC with the external field applied to the tunnel barrier and show the possibility of a transverse (Hall) current in the system.
\begin{figure}[t]
	\includegraphics[width=\columnwidth]{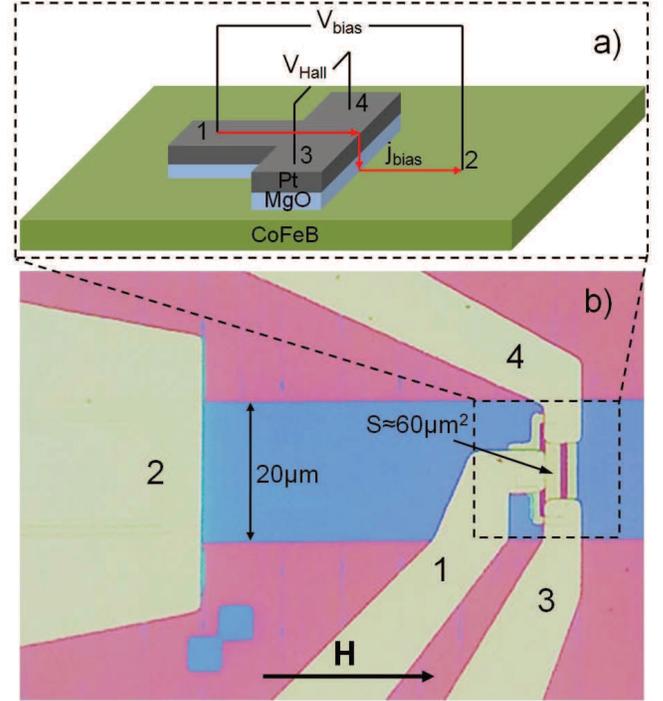}
	\caption{(Color online) (a) Geometry of the sample and measurements. (b) Optical microscope image of the sample. Numbers of the electordes correspond to the scheme (a). ``S'' denotes the area of CoFeB/MgO/Pt sample, the arrow indicates the direction of applied magnetic field.   \label{Fig:Fig.1}}
\end{figure}
\begin{figure*}[t]
	\includegraphics[width=2\columnwidth]{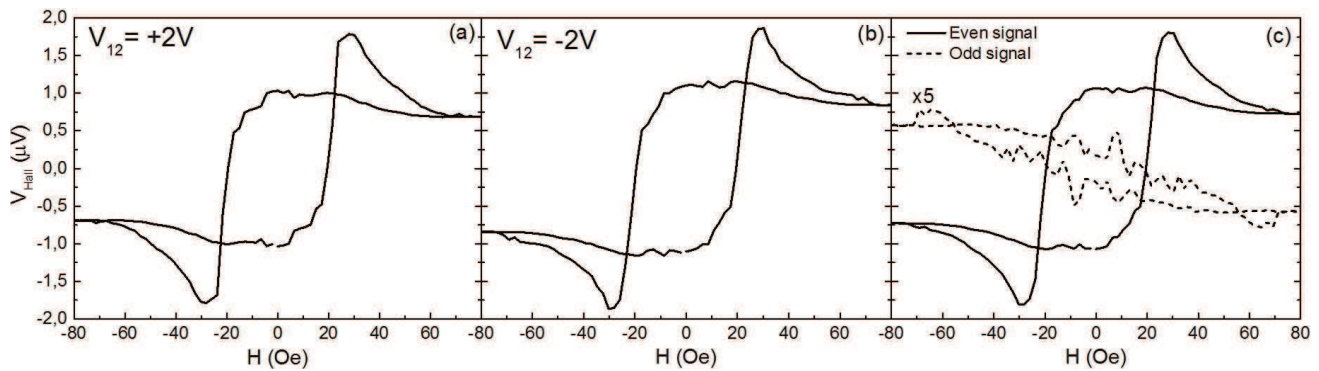}
	\caption{Typical magnetic field dependencies of the Hall voltage for positive (a) and negative (b) DC bias voltage $V_{bias}$ = $\pm$~2~V applied to tunnel barrier. The thickness of the top Pt electrode is 1~nm. (c) – Sum (solid line) and difference (dashed line) of Hall voltage curves (a) and (b) which corresponds to even and odd part of initial Hall signal respectively. \label{Fig:Fig.2}}
\end{figure*}

In our work we experimentally study the THE in a CoFeB/MgO/Pt tunnel junction. Our aim is to find and study spin-orbit effects associated with SOC of the tunneling electron with the external field applied to the tunnel gap. Indeed, the magnitude of the field in the tunnel gap can be as large as $10^9~$V/m, which approaches the magnitude of the internal atomic fields and should lead to the appearance of a transverse current in the system. In order to distinguish this effect from other Hall effects in the system, which are caused by internal atomic electric fields, we hereafter refer to this effect as an extrinsic tunnel Hall effect (eTHE). An important obvious feature of such eTHE should be its quadratic dependence on the applied field. Indeed, in all other cases (AHE, SHE, or intrinsic THE caused by asymmetry of the tunnel barrier) the applied electric field determines only the magnitude of the flowing current in accordance with Ohm’s law. Therefore $U_{Hall}\sim{j}\sim{E}$. In the case of eTHE, the SOC value itself depends on the applied field and $U_{Hall}\sim{jE}\sim{E^2}$.

The studied structures Ta(20)/CoFeB(10)/ MgO(1.5)/Pt(t) (thicknesses are given in nm) are deposited on $\mathrm{Si/SiO_{2}/Si_{3}N_{4}}$ substrates by magnetron sputtering. The base pressure in the chamber does not exceed $10^{-7}$~Torr. To make measurements of the tunneling Hall effect, the Pt and MgO layers are patterned by optical lithography and ion etching into a micron-sized T-shape, as shown in Fig.~\ref{Fig:Fig.1}. Area of the top Pt electrode is about 60~$\mathrm{\mu m^2}$. The bottom CoFeB electrode has rectangular shape of $20\times1000~\mathrm{\mu m}$. It is deposited in an in-plane external magnetic field ($\sim200$~Oe) to induce easy axis anisotropy along the long side of the electrode. The resistance–area product of the tunnel junction is about 15~$\mathrm{k\Omega\cdot\mu m^2}$. To increase the spin polarization of the tunneling electrons, the samples are annealed in vacuum at $330^\circ$~C for 2 hours. The magnetic tunnel junctions CoFeB/MgO/CoFeB fabricated by us using a similar technology as a reference, demonstrates an increase in the tunnel magnetoresistive effect from 10~\% to 200~\% due to thermal annealing. This corresponds to an increase of the spin polarization of the current in the structure from 30 to 70~\% according to the Julier formula~[\onlinecite{Julliere}]. This is due to the crystallization of the initially amorphous CoFeB into the bcc (001) texture from the boundary with MgO (001)~[\onlinecite{Lu}]. The specific crystal structure is essential for the coherent tunneling process and observation of the so-called giant tunnel magnetoresistive effect~[\onlinecite{Yuasa}], which indicates a high spin polarization of current in the structure. The same high spin polarization in the CoFeB/MgO/Pt structure is important because the observed Hall effects directly depend on the spin polarization of tunneling electrons.The thickness of the top Pt electrode is 1, 2 and 10~nm for different samples. In the case of the tunnel Hall effect the Hall current in the normal metal layer (Pt) flows in the subsurface region not exceeding the momentum relaxation length of tunneling electrons which is several nanometers~[\onlinecite{Gall}]. The current caused by the inverse spin Hall effect is induced on the spin relaxation length which varies in the range from 1 to 10 nm according to literature~[\onlinecite{Vlaminck}]. We further show that this length is between 2~nm and 10~nm, from our measurements. Therefore the top Pt electrode must be ultra-thin in order to eliminate the shunting effect and thereby increase the signal-to-noise ratio.

Transport measurements are carried out in the inverse spin Hall effect geometry. A DC bias voltage is applied to leads 1 and 2 ($V_{bias}=V_{12}$), and the Hall signal is measured between leads 3 and 4 depending on the external magnetic field applied in the plane of the film, as shown in Fig.~\ref{Fig:Fig.1}. The magnetization of the CoFeB electrode is aligned along the easy-axis direction defined by the deposition process and can be switched with an external magnetic field $\mathbf{H}$ (Fig.~\ref{Fig:Fig.1}). The Hall effects are measured by applying a bias current from the CoFeB injector into the Pt collector or in the opposite direction. The current is spin polarized according to CoFeB magnetization.

The measured voltage signal $V_{34}$ contains both Hall voltage and additional contribution of magnetoresistance due to asymmetry of the geometric shape of the sample. The Hall effects are odd in the applied magnetic field, while the effects due to the magnetoresistance are even. This makes it possible to extract the Hall effects from the measured signal as an odd part of $V(\mathbf{H})$.
\begin{figure}[t]
	\includegraphics[width=\columnwidth]{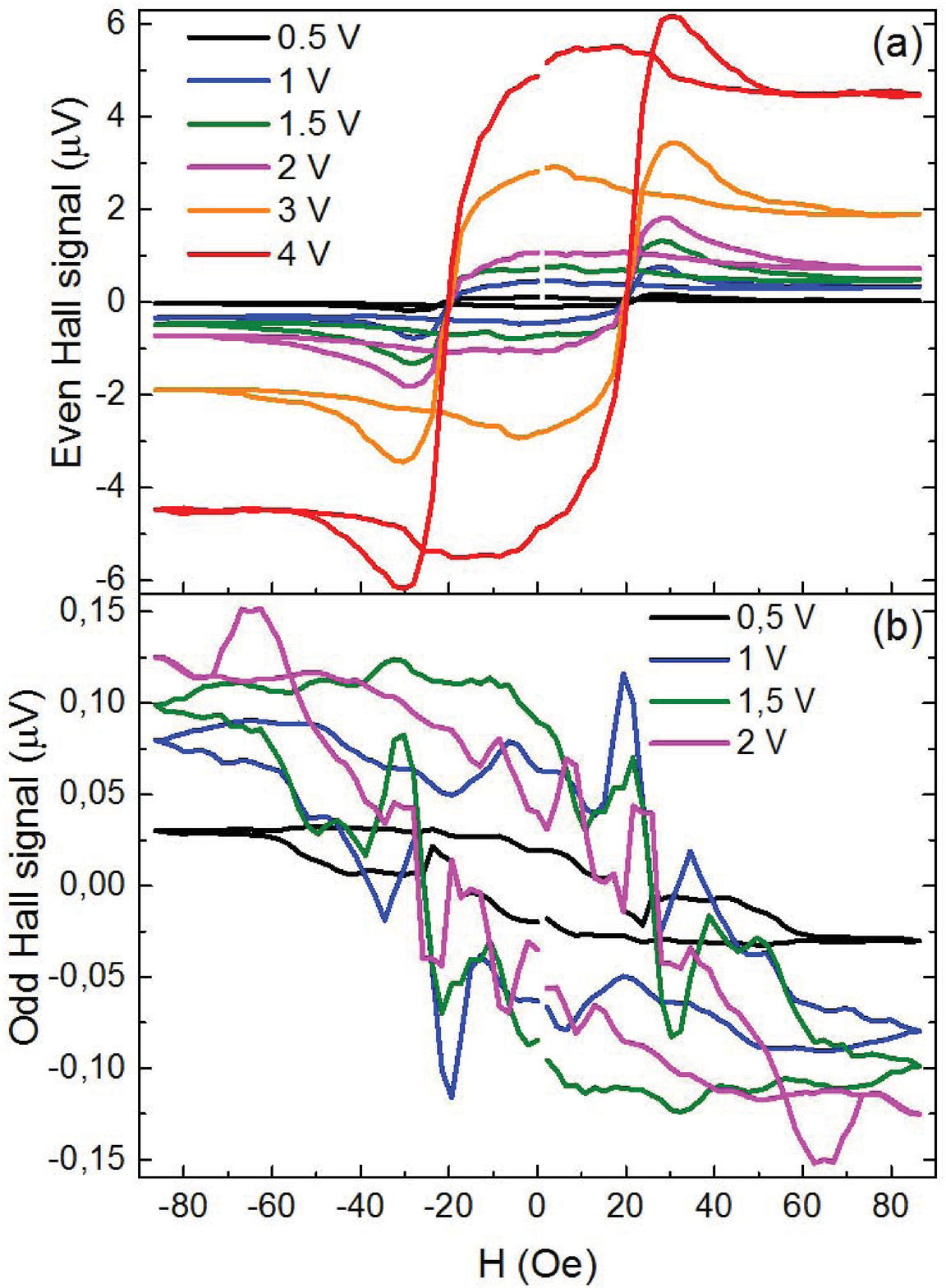}
	\caption{(Color online) (Even $V_{Hall}^{even}$ (a) and odd $V_{Hall}^{odd}$ (b) part of $V_{Hall}(H)$ curves for different values of the applied bias voltage. Thickness of the top Pt electrode is 1~nm.   \label{Fig:Fig.3}}
\end{figure}

The Hall voltage measurements are carried out at different values and both polarities of the DC voltage applied to the tunnel barrier. Fig.~\ref{Fig:Fig.2} (a,b) shows the typical Hall voltage hysteresis. The thickness of the top Pt electrode in this case is 1 nm. The initial signals contain contributions from both the tunnel Hall effect and the anomalous and inverse spin Hall effect. A change in the polarity of the voltage applied to the tunnel contact does not lead to a change in the sign of the measured effect, however, the amplitude of the signal changes. Thus, the signal can be divided into an odd in $V_{bias}$ part which depends on the polarity of $V_{bias}$ and an even part which does not depend on the polarity of the $V_{bias}$, but depends only on its magnitude:
\begin{equation}
\qquad{V_{Hall}^{odd}=\lbrace V_{Hall}(H,+V_{bias})-V_{Hall}(H,-V_{bias})\rbrace/2 \atop V_{Hall}^{even}=\lbrace V_{Hall}(H,+V_{bias})+V_{Hall}(H,-V_{bias})\rbrace/2}
\end{equation}
Both of them are presented in Fig.~\ref{Fig:Fig.2}(c).
\begin{figure}[t]
	\includegraphics[width=\columnwidth]{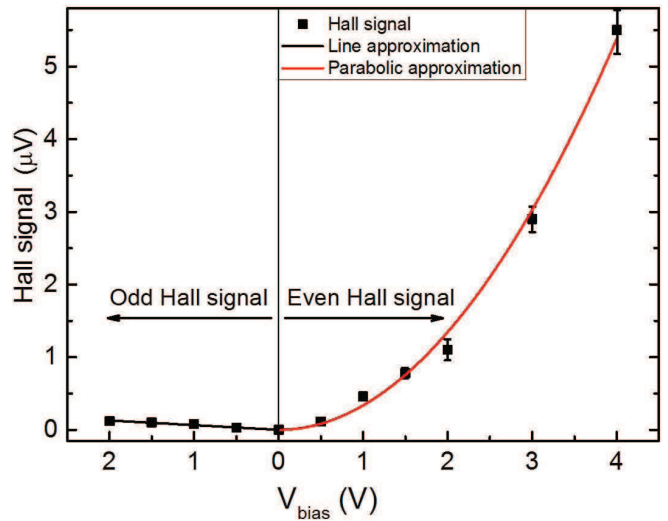}
	\caption{(Color online) Voltage dependence of an even (right side) and an odd (left side) part of initial Hall signal in the sample with 1 nm Pt electrode. An even part is well approximated by a parabola and an odd part approximated by a line.   \label{Fig:Fig.4}}
\end{figure}

Fig.~\ref{Fig:Fig.3} shows the $V_{Hall}(\mathbf{H})$ curves at different values of the voltage $V_{bias}$ applied to the tunnel barrier. The value of the voltage applied directly to the tunnel barrier itself is actually lower than the values of $V_{bias}$ indicated on the graphs, because the resistance of the supply electrodes is approximately two-thirds of the total resistance $R_{12}$. This is established by measuring short-circuit tunnel barriers. It can be seen that an even Hall signal increases non-linearly on the applied electric field (Fig.~\ref{Fig:Fig.3} (a)), which is not observed for an odd signal (Fig.~\ref{Fig:Fig.3} (b)).

Indeed, the dependence of an even Hall signal amplitude on $V_{bias}$ measured in a saturation magnetic field is approximated accurately by a parabola (Fig.~\ref{Fig:Fig.4}). The odd signal is well approximated by a line and probably corresponds to the anomalous and inverse spin Hall effects. So 
\begin{equation}
V_{Hall}=V_{Hall}^{odd}+V_{Hall}^{even}=\alpha V_{bias}+\beta V_{bias}^{2}
\end{equation}
where $\alpha$ and $\beta$ are constants determined from our experiment. The quadratic part represents manifestation of the spin-orbit effect associated with a strong external electric field applied to the tunnel gap. It is the sought-for tunnel Hall effect, predicted in~[\onlinecite{Vedyayev-2}]. The odd part includes all other possible Hall effects, which are linearly proportional to applied voltage, such as an anomalous effect in a magnetic electrode or an inverse spin-hall effect in a platinum electrode. Evidently, they should change sign with the inversion of electric current.

\begin{figure}[t]
	\includegraphics[width=\columnwidth]{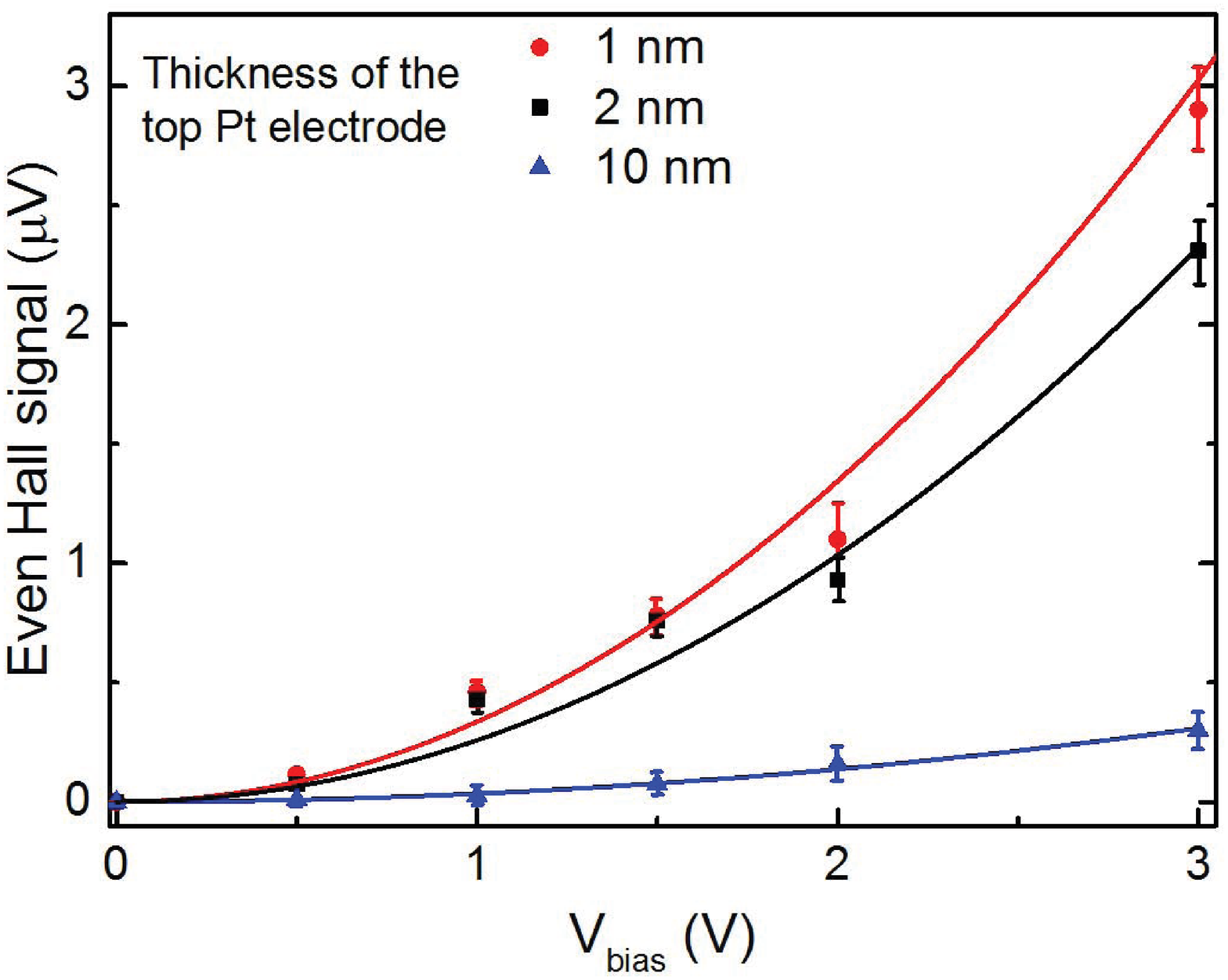}
	\caption{(Color online) Voltage dependence of an even part of the Hall signal for samples with different Pt electrode thickness.   \label{Fig:Fig.5}}
\end{figure}

An increase in the thickness of the top Pt electrode, as expected, leads to a decrease in the Hall voltage and particularly in an even part of initial signal (Fig.~\ref{Fig:Fig.5}).
The data for $\alpha$ and $\beta$ coefficient are presented in the Table \ref{Table1}.

Although the values of $\alpha$ and $\beta$ coefficients are noticeably decreased with increasing thickness of Pt, the general character of dependence (3) is retained.

The linear part of $V_{Hall}$ determined by $\alpha$ allows us to estimate the overall spin Hall angle $\theta_{SH}$ in the investigated system. Using its definition and simple geometric relations one has
\begin{equation}
\theta_{SH}=\frac{1}{P_{s} }\cdot\frac{j_{Hall}^{linear} }{j_{bias}}=\frac{1}{P_{s} }\cdot\frac{R S}{\rho w}\approx 0.008,
\end{equation}
where $R$ is sample resistance (1 $\mathrm{k\Omega}$), $\rho$ is the resistivity of platinum electrode ($1.07\cdot 10^{-7} \mathrm{\Omega\cdot\mu m}$ ), $w$ is the distance between electrodes 3 and 4 ($~8$~$\mathrm{\mu m}$; see Fig. 1), $S$ is the tunnel contact square ($~60$~$\mathrm{\mu m}$) and $P_{s}$ is the spin polarization of the current (70~\%, determined from the data for a reference CoFeB/MgO/CoFeB sample). The spin Hall angle for platinum varies in the range from 0.0067 to 0.027 or even greater, according to literature~[\onlinecite{Vlaminck,Mosendz-1,Mosendz-2}]. Thus we obtain the estimation for our sample which corresponds to literature by the order of value. In addition to this inverse spin Hall effect, we observe a Hall effect caused by spin-orbit scattering of electrons inside the barrier which is an order of magnitude greater.
\begin{table}[t]
\caption{The data for $\alpha$ and $\beta$ coefficients for different Pt electrode thickness\label{Table1}}
\begin{tabular}{ | l | l | l | }
\hline
$d_{Pt}$ & $\alpha$ & $\beta$ \\ \hline
1 & $6.5\cdot 10^{-8} \pm 0.3\cdot 10^{-8}$ & $3.4\cdot 10^{-7} \pm 0.07\cdot 10^{-7}$ \\ \hline
2 & $8\cdot 10^{-8} \pm 10^{-8}$ & $2.6\cdot 10^{-7} \pm 0.4\cdot 10^{-7}$ \\ \hline
10 & $0.6\cdot 10^{-8} \pm 10^{-8}$ & $0.3\cdot 10^{-7} \pm 0.04\cdot 10^{-7}$ \\ \hline
\end{tabular}
\end{table}

It is also possible to estimate the spin-orbit constant $\lambda$ in (1) from our measurements. Using a simple estimation in the limit of small electron tunneling probability we obtain:
\begin{equation}
\lambda\sim\frac{j_{Hall}}{j_{bias}}\frac{L \nu_{f}}{2 P_{s} V_{bias}}\sim 2\cdot 10^{-5}\frac{A\cdot s^{2}}{kg},
\end{equation}
where $L$ is the tunneling barrier thickness, $\nu_f$ is the Fermi velocity, which gives the same result as the calculation~[\onlinecite{Tarasenko}] by the order of value. The estimated constant of spin-orbit coupling is much greater than that constant in vacuum [\onlinecite{Rashba,Petersen}] which leads to an observable Hall effect.

In conclusion, we have investigated transverse transport effect that occurs when spin-polarized electrons tunnel through the tunnel barrier in CoFeB/MgO/Pt junction. We have experimentally found the Hall effect due to spin-orbit coupling of tunneling electrons with a high external electric field applied to the barrier. Tunnel Hall voltage on Pt electrodes depends quadratically on the DC voltage applied to the barrier. This means that it is possible to control and manipulate the spin orbit scattering of spin-polarized electrons in the barrier by voltage.

This research was supported by the Russian Science Foundation (Grant No. 21-12-00271).

\appendix

\end{document}